\documentclass[aps,prb,10pt,superscriptaddress,twocolumn]{revtex4}
\usepackage{graphicx}
\usepackage{amssymb}

\def\cm{cm$^{-1}$}               
\begin{document}

\title{DMTTF-CA revisited: temperature-induced valence and structural instability}

\author{Paolo Ranzieri}\author{Matteo Masino}\author{Alberto Girlando}
\affiliation{Dip. Chimica Generale ed Inorganica, Chimica Analitica e 
Chimica Fisica, Universit\`a di Parma, Parco Area delle Scienze,
43100-I Parma, Italy}
\author{Marie-H\'el\`ene Lem\'ee-Cailleau}
\affiliation{Institut Laue-Langevin, 38042 Grenoble, France}
\begin{abstract}
We report a detailed spectroscopic investigation of temperature-induced valence
and structural instability of the mixed-stack organic charge-transfer (CT) crystal
4,4'-dimethyltetrathiafulvalene-chloranil (DMTTF-CA). DMTTF-CA is a derivative of
tetrathiafulvalene-chloranil (TTF-CA), the first CT crystal exhibiting
the neutral-ionic transition by lowering temperature.
We confirm that DMTTF-CA undergoes a continuous variation of the ionicity on going from room temperature down to $\sim$ 20 K, but remains on the neutral
side throughout. The stack dimerization and cell doubling,
occurring at 65 K, appear to be
the driving forces of the transition and of the valence instability. 
In a small temperature interval just below the phase transition
we detect the coexistence of molecular species with slightly different
ionicities. The Peierls mode(s) precursors of the stack dimerization
are identified.
\end{abstract}

\maketitle

\section{Introduction}
\label{sec:intro}
Organic charge-transfer (CT) crystals made up by $\pi$ electron-donor
(D) and electron acceptor (A) molecules often exhibit a typical stack
structure, with D and A molecules alternating along one direction.\cite{herbestein71,soos76} The quasi-one-dimensional
electronic structure is stabilized by the CT interaction between
D and A, so that the ground state average charge on the molecular 
sites, or degree of ionicity, $\varrho$, assumes values between
0 and 1. Crystals characterized by $\varrho \lesssim$ 0.5 are
\textit{conventionally} classified as quasi-neutral (N),
as opposed to the quasi-ionic (I) ones, with $\varrho\gtrsim$ 0.5.
As discussed for the prototypical
system of tetrathiafulvalene-chloranil (TTF-CA),\cite{torrance}
a few CT salts have N-I and
Peierls transition, in which $\varrho$
changes rapidly and the regular stack dimerizes,
yielding a potentially ferroelectric ground state.\cite{girlando04}
N-I transitions are valence instabilities implying a \textit{collective}
CT between D and A sites, and as such are accompanied by many
intriguing phenomena, such as dielectric constant anomalies, current-induced
resistance switching, relaxor ferroelectricity, and so on.\cite{horiuchirev}
The isostructural series formed by 4,4'-dimethyltetrathiafulvalene 
(DMTTF) with substituted CAs, in which one or more chlorine
atom is replaced by a bromine atom, is particularly interesting. In this case,
in fact,
the transition temperature and related
anomalies can be lowered towards zero by chemical or physical
pressure, attaining the conditions of a quantum
phase transition.\cite{horiuchi01,horiuchi03,okimoto05}

Albeit several aspects of the N-I transition in Br substituted DMTTF-CA
family are worth further studies, the motivation of the present work
is far more limited, as we want first of all clarify the mechanism of the transition in the pristine compound, DMTTF-CA. Despite intensive
studies,\cite{horiuchi01,horiuchi03,okimoto05,aoki93,collet01,collet02}  the transition still presents controversial aspects.
Through visible reflectance spectra of single crystals and
absorption spectra of the powders, Aoki\cite{aoki93} suggested that by
lowering the temperature below 65 K, DMTTF-CA does not
undergo a full N-I transition, but forms a phase in which both
N ($\varrho = 0.3-0.4$) and I ($\varrho = 0.6-0.7$) species are
present. The structural investigation as a function of 
temperature\cite{collet01} put in evidence a fundamental aspect
of the transition, only implicit in Aoki's work:\cite{aoki93}~
At 65 K the unit cell doubles along the \textit{c}
axis (\textit{a} is the stack axis). The order parameter of the transition, which is second-order, is the cell doubling coupled with the dimerization.\cite{collet01} 
So above 65 K the cell contains one stack, and at 40 K contains two
stacks, both dimerized, and inequivalent (space group $P1$). From the
bond distances, $\varrho$ is estimated at 0.3 and 0.7-0.8
for the two stacks, respectively.\cite{collet01} In this view, and
considering that
the two stacks are dimerized in anti-phase, at low temperature DMTTF-CA has a
\textit{ferrielectric} ground state.

However, the above scenario has been questioned.\cite{horiuchi01,horiuchi03}  Polarized single crystal infrared (IR) reflectance measurements 
suggests that N and I stacks do not cohexist. Only one ionicity is observed,
changing continuously
from about 0.25 at room temperature to about 0.48 at 10 K,
the maximum slope in the $\varrho(T)$ occurring around 65 K.
The crystal structure at 14 K indicates a $P\bar 1$ space group, with two
equivalent, dimerized stacks in the unit cell, and 
\textit{anti-ferroelectric} ground state.\cite{horiuchi03}
According to this picture, the mechanism of DMTTF-CA phase transition
is very similar to the other N-I transitions.\cite{girlando04,horiuchirev}
The Madelung energy change yields an appreciable change of $\varrho$
(about 0.1) within a few
degrees of temperature, accompanied by a stack dimerization.
The cell doubling appears to be a secondary aspect, whereas the
most important feature is the continuous variation of $\varrho$,
as opposed for instance to the discontinuous, first order transition of
TTF-CA.\cite{girlando04}

Some questions remain however unanswered in the above
picture.\cite{horiuchi01,horiuchi03} The transition displays a continuous
ionicity change with $T$, and consequently one would expect huge anomalies at the transition, whereas for instance the dielectric constant
increase at $T_c$ is less than in the case of TTF-CA.\cite{girlando04,horiuchirev} Furthermore, what is
the driving force of the transition? In TTF-CA, the N-I transition
is attributed to the increase of Madelung energy by the lattice
contraction.\cite{torrance}  If it is so also for DMTTF-CA, what is the role
of cell doubling? Finally, although $P1$ and $P\bar 1$
space groups are sometimes difficult to disentangle by X-ray diffraction, the issue
of the different published structures is not solved, both
exhibiting good confidence factors in the refinement process \cite{horiuchi03,collet01}

In order to clarify these open questions, and to understand
the mechanism of the phase transition in DMTTF-CA, we have
decided to collect and re-analyze complete polarized IR and Raman
spectra of DMTTF-CA single crystals, along the same lines followed
for TTF-CA.\cite{girlando83,masino03,masino06} Indeed, a careful analysis
can give information about $\varrho$, stack
dimerization, and the Peierls mode(s) inducing it.
Vibrational spectra give information about the \textit{local}
structure, and from this point of view are complementary
to the X-ray analysis, which probes long range order.
We shall show that DMTTF-CA transition can hardly be classified
as a N-I transition, the most important aspect being the stack
dimerization and cell doubling. We shall also offer
some clues about the origin of the discrepancies in the two
X-ray determinations.\cite{horiuchi03,collet01}

\section{Experimental}
\label{sec:expt}

DMTTF-CA single crystals have been prepared as previously described.\cite{collet01} 
The IR spectra (600-8000 \cm) have obtained with a Bruker IFS66 FTIR spectrometer,
equipped with A590 microscope.
Raman spectra have been recorded
with a Renishaw 1000 micro-spectrometer.
The excitation of Raman has been achieved with a Lexel Krypton laser
($\lambda$ = 647.1 nm), backscattering geometry, with less than 1 mW power
to avoid sample heating.
A pre-monochromator has been used for the low-frequency spectra
(below 200 \cm).
For the high frequency Raman spectra
we report only the spectra obtained for incident and scattered light
both polarized perpendicularly to the stack axis,
($\perp \perp$) in the conventional notation. In this arrangment the in-plane molecular modes, notably the totally symmetric ones, are more clearly visible.
The spectral resolution of IR and Raman spectra is 2 \cm.

Temperatures down to 10 K have
been reached with a ARS closed-circle cryostat, fitted under
the IR and Raman microscopes. The temperature reading on the cold finger
has been tested and considered accurate to $\pm$2 K for the Raman and
IR reflectance measurement, where silver paste has been used
to glue the sample to the cold finger. For IR absorption the temperature reading
is far less accurate, due to the unperfect thermal contact between the
sample and the KBr window on the cold finger. Temperature reading corrections
have been applied based on the comparison with the reflectivity data.
DMTTF-CA reflectivity has been normalized to that of an Al mirror,
without further corrections. Therefore the reflectance values  
are not absolute, and relative values can be compared with confidence
only within each low-temperature run. 
We consider reflectance values
of the spectra below 20 K not reliable in any case, because the deposition
of an unknown contaminant on the DMTTF-CA surface introduces high
noise above $\sim$ 2000 \cm.

\section{Results}
\subsection{Valence instability}
\label{sec:valence}

\begin{figure}
\begin{center}
       \includegraphics*[scale=0.54] {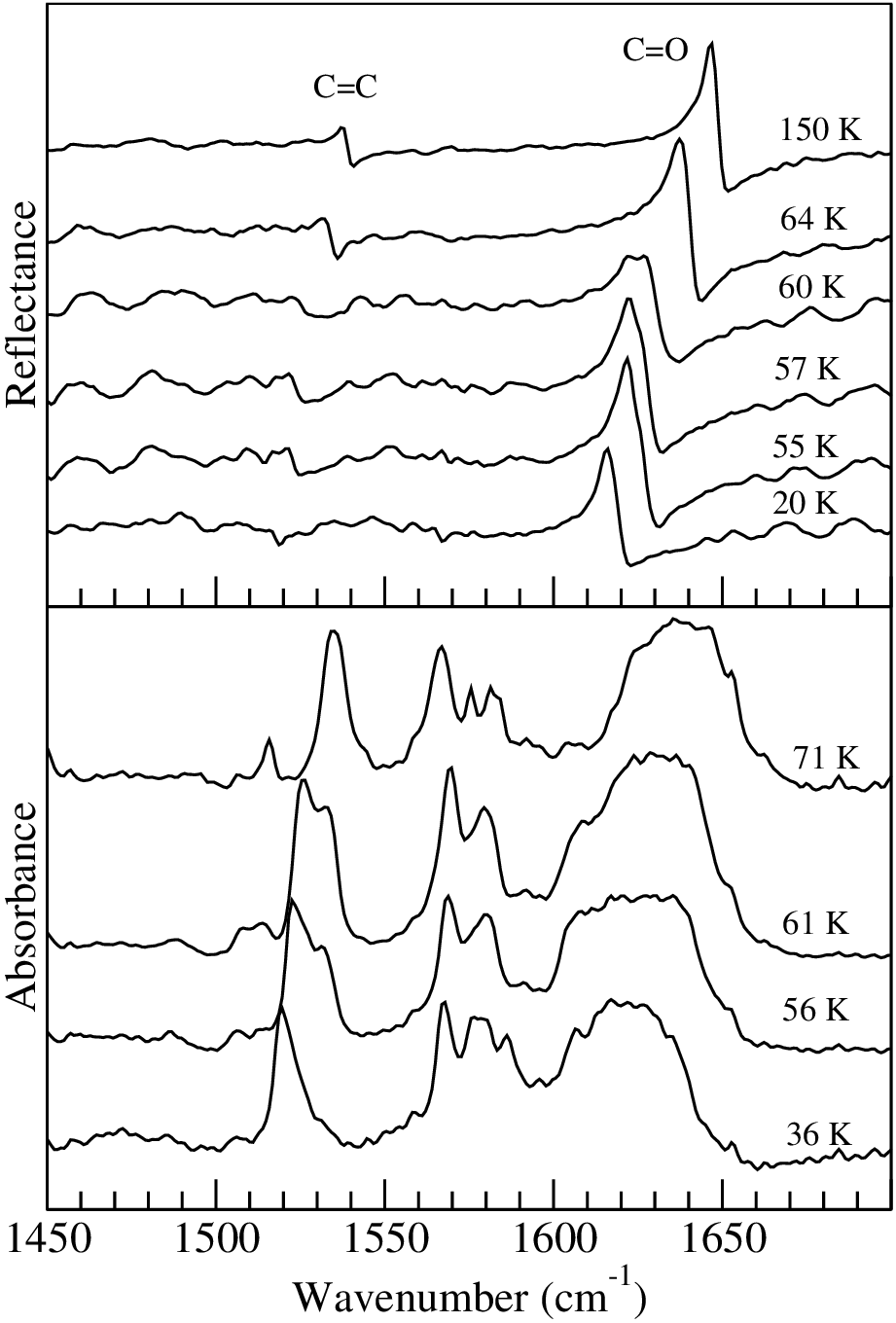}
\caption {Temperature evolution of DMTTF-CA reflectance and absorbance spectra,
polarized perpendicularly to the stack. The bands corresponding to
C=O and C=C stretchings of CA are marked.}
       \label{fig:IRperp}
\end{center}
\end{figure}

The first question we address is that of the ionicity as a function of temperature.
To such aim, we have collected  both IR reflectance and absorbance spectra, with
polarization perpendicular to the stack axis. The two types of spectra allow us
to ascertain whether probing the surface or the bulk yields the same result.
Unfortunately, we were unable to obtain crystals sufficiently thin to avoid
saturation of the most intense absorption bands,
so the information provided by the two types of spectra are complementary.
Fig.~\ref{fig:IRperp} shows some examples of spectra
as a function of temperature in the frequency range 1500-1700 \cm.
The two structures at 1649 \cm  and 1539 \cm (at 150 K) are assigned
to the $b_{1u} \nu_{10}$ and $b_{2u} \nu_{18}$ modes
of the CA moiety, corresponding to the C=O and C=C antisymmetric
stretching vibrations, respectively.\cite{girlando83}
The C=O stretching mode is the most sensitive to the molecular charge,
so it has been almost invariably used to estimate the ionicity.
However, recent investigations on  CA and CA$^-$ molecular vibrations
have shown that also the C=C mode should be a good $\varrho$
indicator.\cite{katan96,ranzieri} Therefore we shall use C=O
$b_{1u} \nu_{10}$ as a primary $\varrho$ indicator, and C=C
$b_{2u} \nu_{18}$ mode as secondary, internal consistency probe.

\begin{figure}
 \centering
 \begin{center}
 \includegraphics[scale=0.42]{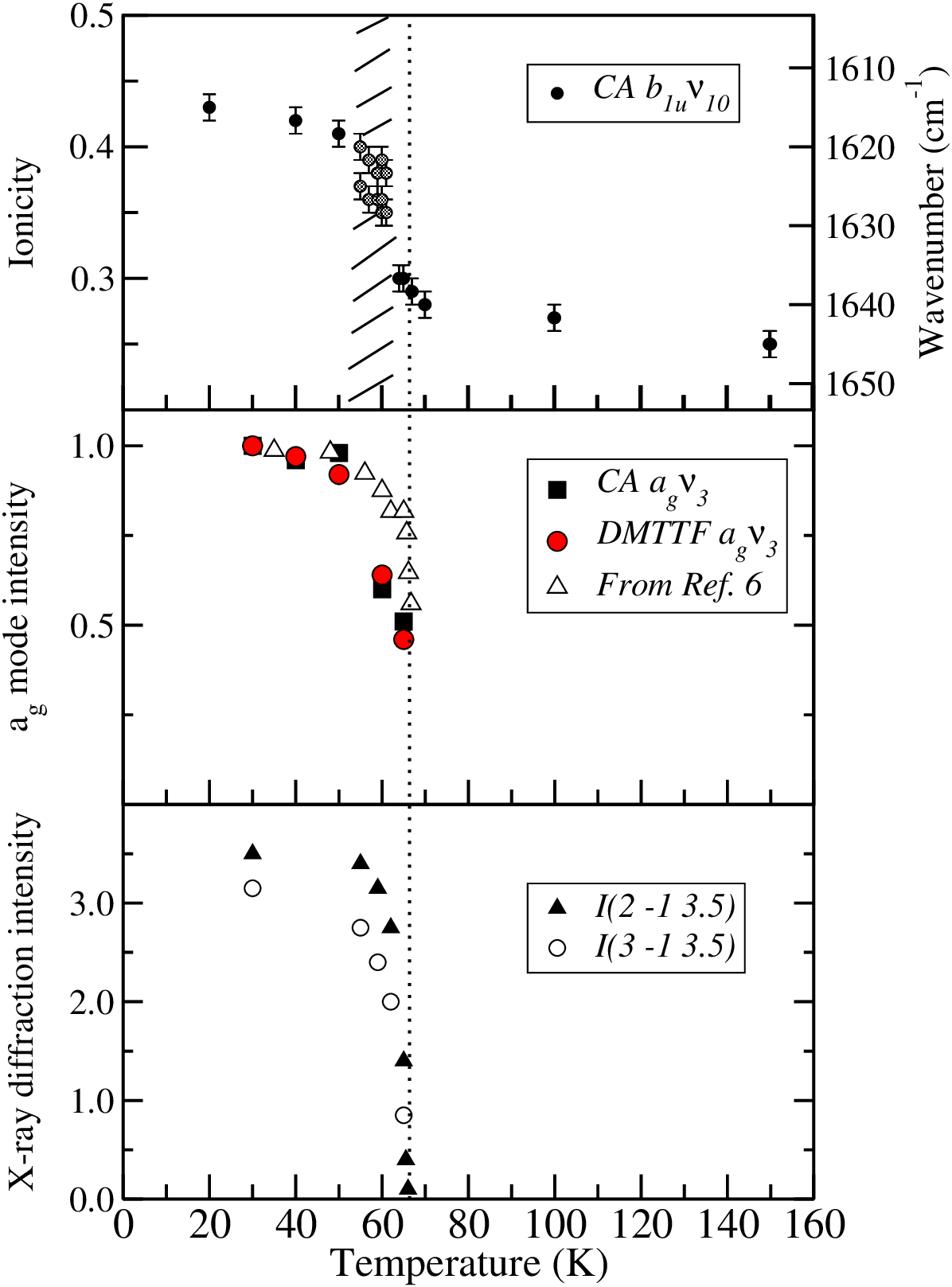}
\end{center}

 \caption{Temperature evolution of three DMTTF-CA observables.
Top panel: Ionicity $\rho$. Middle Panel: Normalized intensity of IR vibronic bands,
connected to the stack dimerization amplitude. Bottom panel: Intensity 
of X-ray reflections signaling the cell doubling (from
Ref.~\onlinecite{collet01}). The vertical dashed line marks
the critical temperature $T_c \simeq$ 65 K. }
 \label{fig:ionint}
\end{figure}

As the C=O stretching mode saturates in absorption (Fig.~\ref{fig:IRperp},
bottom panel), we have performed
the usual Kramers-Kronig transformation of the reflectance
spectra. From the frequency reading of the C=O $b_{1u} \nu_{10}$ mode we
have estimated the ionicity by the usual relationship:
$\varrho = [\bar\nu(0)-\bar\nu(\varrho)]/\Delta_{ion}$
where $\bar\nu(0)$ is the C=O stretching frequency of the neutral molecule
and $\Delta_{ion}$ is the ionization frequency shift.\cite{girlando83}
The resulting $\varrho(T)$ is reported in the top panel
of Fig.~\ref{fig:ionint}.

The top panel of Fig.~\ref{fig:ionint} is rather similar
to the corresponding one of Ref. \onlinecite{horiuchi01}.
Indeed, our spectra (Fig. ~\ref{fig:IRperp}) do not show
the onset of a strong band around 1580 \cm~ below 65 K,
that according to Aoki \textit{et al.}\cite{aoki93} 
signals the presence of $\varrho \sim 0.7$ species. 
Following Ref. \onlinecite{horiuchi01}, we attribute
the band around 1580 \cm ~to an activated
$a_g$ mode, present in the spectra polarized parallel
to the stack (see Section \ref{sec:dimerization}). However, our
data show two significant differences compared with
Horiuchi \textit{et al.} results.\cite{horiuchi01}

First of all, our $\varrho(T)$ curve is consistently shifted
downward by about 0.05 $\varrho$ units with respect to the
corresponding one of Ref. \onlinecite{horiuchi01}. As a consequence,
the maximum ionicity at the lowest temperature is well below the N-I
borderline, namely 0.43 instead of 0.48.
We believe the discrepancy is due to the extrapolation
involved in the Kramers-Kronig transformation from reflectance
to conductivity. We have indeed verified that different extrapolation procedures
may yield different frequencies, and we have chosen  extrapolation parameters 
giving conductivity spectra with frequencies matching those read in absorption
(bottom panel of Fig.~ \ref{fig:IRperp}).
Our datum is also in agreement with what reported by Aoki,\cite{aoki93}
who assigned a $\varrho$ value of $\sim 0.3-0.4$ to the
N species at 20 K (although, as
stated above, we do not see the insurgence of bands due to I
species).

The second important difference of our results compared with those of
Horiuchi \textit{et al.} \cite{horiuchi01} is that just below the phase transition temperature, between 62 and 54 K, the C=O stretching
mode  shows a clear doublet structure (Fig. ~\ref{fig:IRperp}, top panel), suggesting the presence of two slightly
differently charged molecular species. The indication is confirmed
by the band due to the C=C stretching mode, which also shows a doublet structure,
clearly seen in absorption (bottom panel of Fig.~\ref{fig:IRperp}).
The frequency (and ionicity) difference is small, but clearly
visible irrespectively of the direction of temperature change,
and reproducible in different runs. Actually, a hint of a doublet
structure is visible also in Ref. \onlinecite{horiuchi01} spectra,
but it was interpreted as a band broadening. In Fig. ~\ref{fig:IRperp}
the dashed area indicates the temperature interval of this coexistence.

To summarize, our results present a valence instability scenario
different from both the previously reported ones.\cite{aoki93,horiuchi01}
The ionicity change appears to be continuous across the phase transition.
The crystal remains
neutral ($\varrho \sim 0.43$ at 20 K), therefore excluding the simple term 
of N-I transition: it is better to refer to it as a valence instability.
Finally, in a temperature interval of less than 10 K below 65 K there is
coexistence of two species with slightly different molecular ionicity,
both on the neutral side, with $\varrho \sim$ 0.36 and 0.38.

\subsection{Dimerization instability}
\label{sec:dimerization}

\begin{figure}[htp]
 \centering
 \begin{center}
 \includegraphics[scale=0.37]{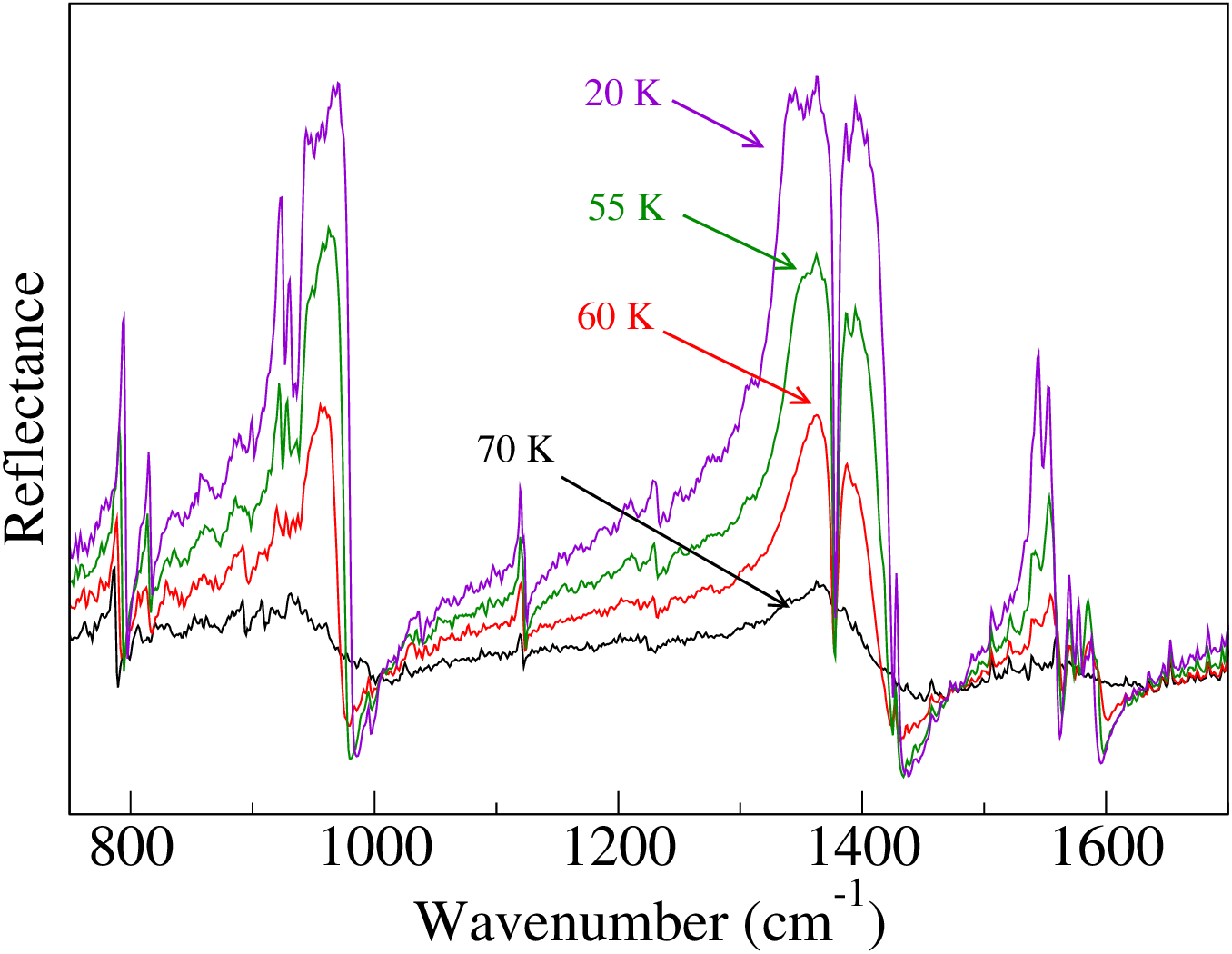}
\end{center}
\caption{Temperature dependence of DMTTF-CA reflectance spectra polarized along the stack. The reflectance scale is not reported, as
only relative reflectance values can be trusted (see Experimental).}
\label{fig:reflpar}
\end{figure}

It is well known that IR spectra polarized along the stack are sensible to the
symmetry breaking associated with stack distortions. In fact, the loss of
inversion center on the molecular units makes the Raman-active totally-symmetric ($a_g$)
molecular modes also IR active, with huge intensity due to their coupling
with the CT electronic transition (IR ``vibronic bands'').\cite{girlando83}
In addition, it has been recently shown that the IR spectra polarized
parallel to the stack, associated with Raman, also yield information about pre-transitional phenomena, like the softening of the phonons inducing the stack distortion.\cite{masino03} To investigate these aspects of the DMTTF-CA phase transition,
we have collected the IR reflection spectra polarized parallel to
the  stack axis. 
Fig.~\ref{fig:reflpar} shows some examples of the spectra at different
temperatures across the phase transition.
Fig. \ref{fig:reflpar} shows the insurgence of strong features below $T_c$.
The spectra are identical to those of Ref. \onlinecite{horiuchi01}, and
we follow a similar analysis. We focus the attention on the two main
features around 1400 and 980 \cm, associated with the $a_g \nu_3$ mode
of the TTF skeleton in the DMTTF moiety, and to the
$a_g\nu_3$ mode of CA, respectively.\cite{girlando83}
Their intensity $f_j$ as a function of $T$ is extracted 
by fitting the reflectance spectra
with a Drude-Lorentz oscillator model, where the
dielectric constant is given by:
\begin{equation}
\epsilon(\omega) = \epsilon_\infty + \sum_j \frac{f_j}
{\omega_j^2-\omega^2 -i\omega \gamma_j} 
\end{equation} 
In this equation $\epsilon_\infty$ is the high-frequency dielectric
constant, and $\gamma_j$ is the line width.
Now, $f_j$ is related to $\delta$, the dimerization
amplitude.\cite{painelli03} The relationship is not of direct
proportionality, but in any case $f_j$ gives an indication of the
increase of the dimerization amplitude at the
transition.\cite{painelli03} The $f_j(T)$ for the
two modes, normalized
at the lowest temperature value (20 K), are reported in the middle panel
of Fig.~\ref{fig:ionint} for temperatures below 65 K.
Indeed, around the critical temperature
one cannot disentangle the contribution to the intensity
coming from the combination modes, as discussed in detail in the next Section.
In the same panel we report, for comparison,
the normalized intensity relevant to
the DMTTF $a_g$ mode, as given in Ref. \onlinecite{horiuchi01}
for the same temperature range.

The middle panel of Fig. \ref{fig:ionint} shows that
the IR oscillator strengths, related to the stack dimerization
amplitude $\delta$,\cite{painelli03} display a
behavior typical of an order parameter relevant to a second-order phase
transition. It is instructive from this point of view to compare the present
data with the intensity of the X-ray diffraction spots
related to the cell doubling, as reported in Ref. \onlinecite{collet01}
and shown in the bottom panel of Fig.~\ref{fig:ionint}. 
The two sets of data exhibit the same behavior, demonstrating
that the cell doubling, as detected by X-ray, and stack
dimerization amplitude, as detected by IR,
occur at the same time,
representing two inseparable aspects of 
DMTTF-CA phase transition. On the other hand, the
comparison of the three panels in Fig.~\ref{fig:ionint}
puts in evidence that the valence instability actually
\textit{follows} the structural modification. Dimerization
and cell doubling start at $T_c \simeq$  65 K, whereas the
rapid increase in $\varrho$ occurs slightly below, and
implies the simultaneous presence of species
with two slightly different ionicities in a $T$
interval of about 10 degrees.

\subsection{Lattice phonons}
\label{sec:lattice}

If DMTTF-CA phase transition is displacive, it should
imply the occurrence of soft phonon(s) yielding the 
stack dimerization and the cell doubling.
In a simplified but effective view,
we can think of the phase transition as due to just one critical
phonon, with wavevector $c^*/2$. 
Such a phonon belongs to a phonon branch
that corresponds to stack dimerization
along the $a$ crystal axis, and at the zone-center
is optically active.
The driving force of the transition is then provided by the
Peierls mechanism, which couples the zone-center
dimerization mode, i.e., the Peierls mode,
with the CT electronic structure.\cite{soos04}
The electron-phonon causes the softening of the Peierls
mode, eventually laeding to stack dimerization.
However, in the proximity of the phase transition, where
interstack interactions are more effective,\cite{collet02}
the Peierls mode evolves to a stack dimerization out-of-phase
in nearest-neighbors cells, when it softens
yielding the cell doubling along the crystallographic direction $c$.
Of course, in the complicated phonon structure of a molecular
crystal like DMTTF-CA, the Peierls mode may result from the superposition (mixing) of several phonons, all directed along the stack.
A spectroscopic investigation of DMTTF-CA, along the lines
already developed for TTF-CA,\cite{masino03,masino06} should yield
the identification of these phonons or of the resulting ``effective''
Peierls mode.

Phonons coupled to the CT electrons along the chain are
most likely inter-molecular, or lattice, phonons.\cite{masino06}
We start by classifying the lattice phonons and their
Raman and IR activity by adopting the rigid molecule approximation.
This approximation is known to be not fully valid for TTF-CA,\cite{masino06}
but is the only reasonable starting point in the lack of explicit calculations
of DMTTF-CA phonon dynamics. 
Then, in the high temperature (HT) phase ($P\bar1, Z=1$)\cite{collet01} we expect 9 opticallly active lattice modes,
$6A_g(\mathcal{R}) + 3A_u(\mathcal{T})$. The Raman active $A_g$ modes can be described
as molecular librations ($\mathcal{R}$), and are decoupled from the CT electrons.
Coupling is instead possible for the IR active $A_u$ phonons,
which indeed correspond to translations ($\mathcal{T}$).  There are
no symmetry constraint about the direction of molecular displacements,
so we may have some component of all the three $A_u$ phonons along
the stack axis, contributing to the Peierls mode.

In the low-temperature (LT) phase it is
not clear if the two stacks inside the unit cell are
inequivalent, with space group $P1$,\cite{collet01}
or equivalent, with space group $P\bar 1$.\cite{horiuchi03}
Since in any case the inequivalence is small, for the spectral predictions
we find more convenient to use the centro-symmetric description.
The center of inversion is between the two
stacks in the unit cell, then we expect 21 optically
active modes, $12 A_g(6\mathcal{T} + 6\mathcal{R})$ and $9 A_u (6\mathcal{R}+
3\mathcal{T})$. Therefore, the phonons modulating the CT integral
are IR active in both HT and LT phases, whereas
the cell doubling in the LT
phase makes
Raman active 6 translational
phonons, which correspond to the coupled
in-phase displacements of the two chains.

\begin{figure}
\begin{center}
       \includegraphics*[scale=0.56] {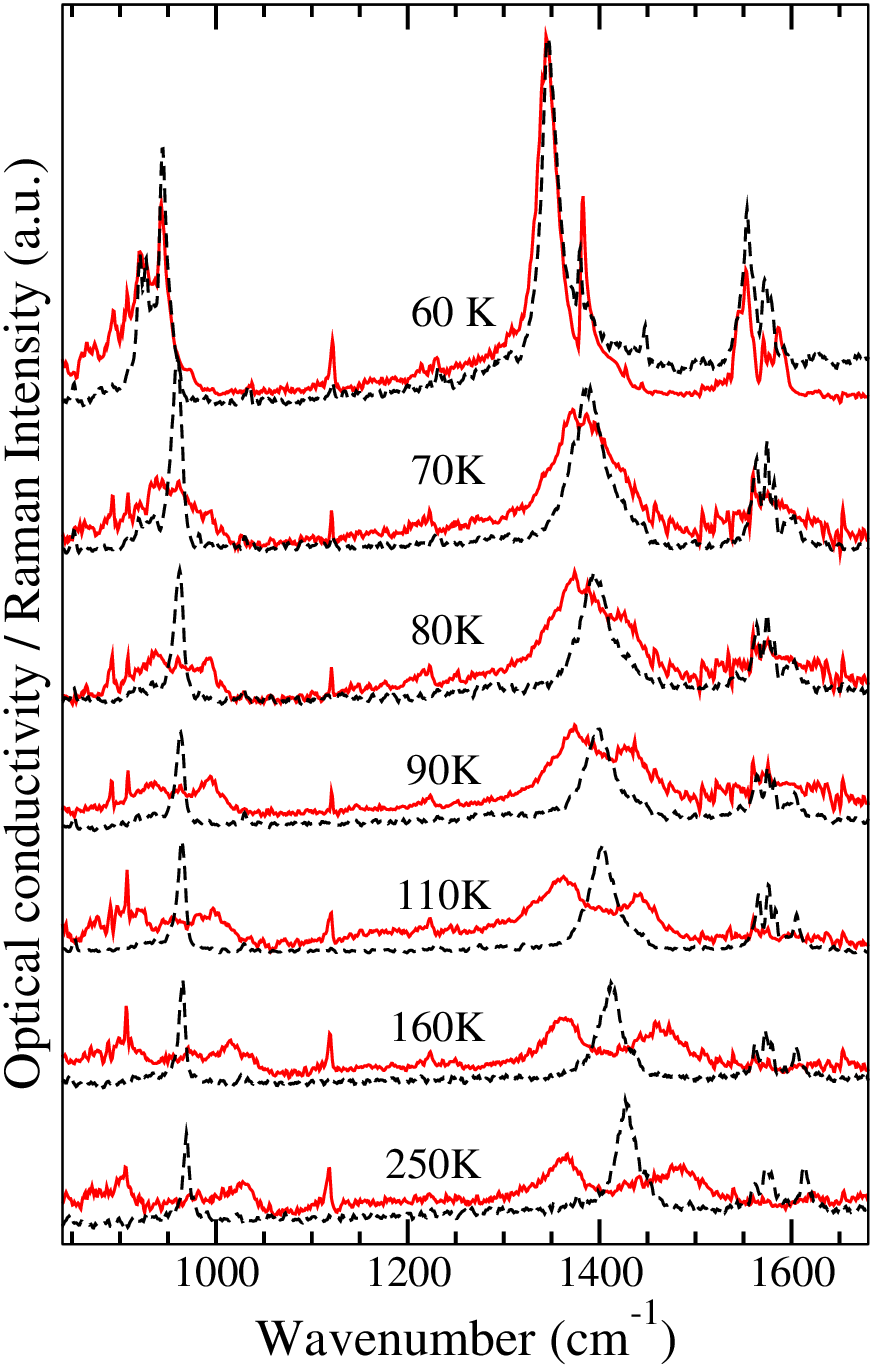}
       \caption {Temperature dependence of DMTTF-CA Raman spectra (black,
        dashed line), and IR conductivity spectra polarized parallel to
	the stack axis (red, continuous line)}
       \label{fig:RamanIR}
\end{center}
\end{figure}

Direct investigation of Peierls coupled modes in the far-IR
(5-200 \cm) is not an easy task.\cite{masino06} However, 
in the case of TTF-CA it has been shown\cite{masino03} that useful information can be obtained from the comparison between Raman and IR spectra polarized parallel to the stack, in the frequency region of molecular vibrations.
We have then performed the Kramers-Kronig transformation of the reflectivity
data of Section \ref{sec:dimerization} (Fig. \ref{fig:reflpar}),
obtaining the optical conductivity spectra which are compared with Raman
in Fig. \ref{fig:RamanIR}.

\begin{figure}
\begin{center}
       \includegraphics*[scale=0.49] {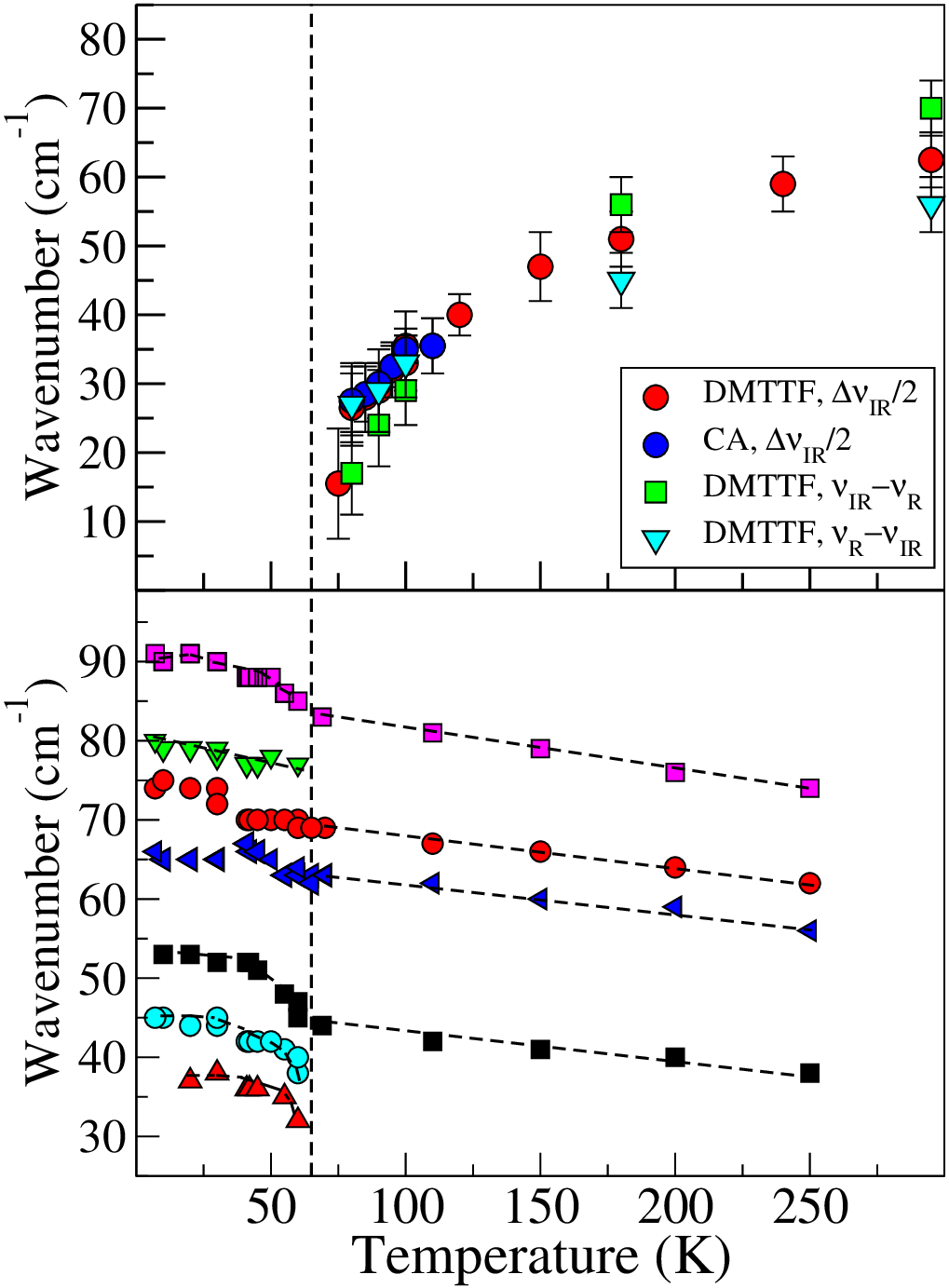}
       \caption {Top panel: Frequency difference between Raman DMTTF
         $a_g \nu_3$ and CA $a_g \nu_3$ bands and the corresponding
         IR sidebands of Fig.~\ref{fig:RamanIR}
as a function of temperature. Bottom panel: Temperature evolution of the
most intense low-frequency Raman bands. The vertical dashed line marks the
critical temperature.}
       \label{fig:peierls}
\end{center}
\end{figure}

We again focus attention on the $a_g\nu_3$ mode of the TTF skeleton in the
DMTTF moiety and to the $a_g\nu_3$ mode of CA, which correspond to the two
most prominent Raman bands of Fig. \ref{fig:RamanIR}, located around 1400 \cm and 980 \cm, respectively.\cite{girlando83} The IR spectra above 80 K
exhibit pairs of absorptions (``side-bands''), symmetrically located above and below
the just mentioned Raman bands. The side-bands are quite naturally interpreted
as sum and difference combination bands between the corresponding $a_g$
mode and a lattice phonon.\cite{masino03} By lowering temperature the
side-bands approach each other, and around 80 K they start to coalesce and
to overlap to the central Raman band. On the other hand, below the transition
temperature (for instance, 60 K in Fig.~\ref{fig:RamanIR}) there is coincidence
between Raman and IR bands. As discussed in Section \ref{sec:dimerization},
both are indeed due to the same $a_g$ molecular
vibration, active in both type of spectra due to the symmetry breaking
connected to the stack dimerization.

Analysis of the side-bands therefore gives information on
the Peierls mode in the HT phase.\cite{masino03}
In the top panel of Fig.~\ref{fig:peierls} the frequency difference
between the DMTTF $a_g \nu_3$ Raman band and
the corresponding IR side-bands is plotted as a function of temperature.
We also plot the frequency semi-difference between the side-bands
associated with both DMTTF and CA $a_g \nu_3$ modes.
Fig.~\ref{fig:peierls} shows that the data points coincide
within experimental error, supporting the idea that \textit{the same}
lattice mode is involved in the combination, and clearly indicates a soft mode behaviour.
This softening suggests that this lattice
phonon is indeed the Peierls mode, or to be precise, the
``effective'' Peierls mode, resulting from the superposition
of several modes coupled to the CT. We cannot follow the
frequency evolution down to the transition temperature, since below 80-75 K
it becomes impossible to separate the contribution of the two
side-bands, letting aside the interference from the fluctuations
occurring near the phase transition.

\begin{figure}[ht]
\begin{center}
       \includegraphics*[scale=0.45] {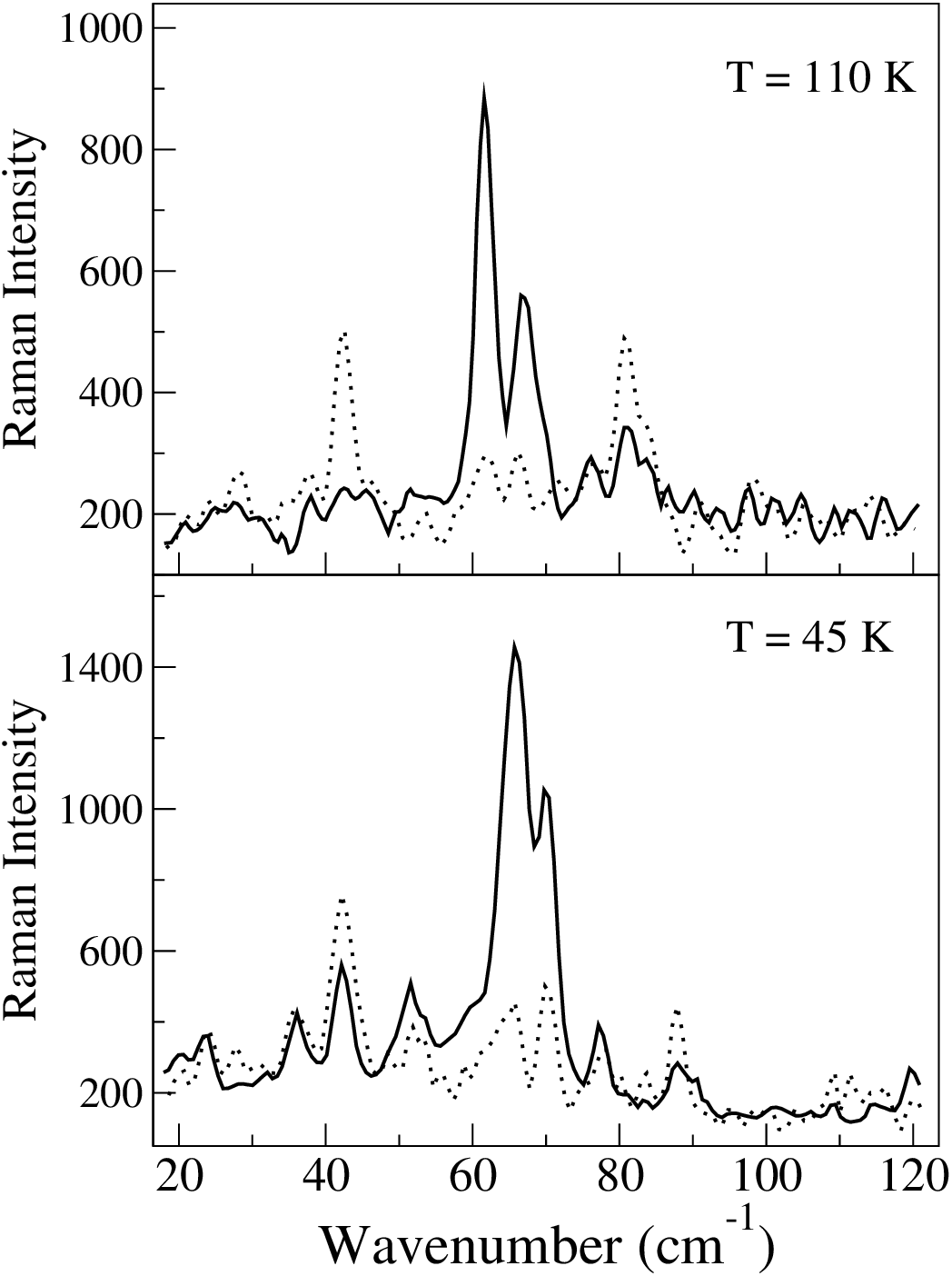}
       \caption {Top panel: polarized low-frequency Raman spectra
 of DMTTF-CA at 110 K (HT phase). Bottom panel: polarized
 low-frequency Raman spectra of DMTTF-CA at 45 K (LT phase). Continuous
 line indicates ($\perp \perp)$ polarization; dashed line 
($\parallel \perp$) polarization (see text).
                  }
       \label{fig:Rlattice}
\end{center}
\end{figure}

We now turn attention to DMTTF-CA LT phase, 
where the Peierls
mode(s) are active in Raman in addition to IR. We have
then measured the Raman spectra in
the low-frequency region in order to identify 
the possible soft phonons. Low-frequency Raman also gives indications about the cell doubling, given the difference in the number of phonons present in the
two phases.

An example of the low-frequency Raman spectra above and below
the phase transition is shown Fig.~\ref{fig:Rlattice}.
We report the spectra with two polarizations. In one, both incident and scattered light are polarized perpendicularly 
to the stack axis  ($\perp \perp$ spectra). In the
other, the polarization of the incident light is
rotated parallel to the stack axis, the polarization
of the scattered light being kept perpendicular to it
($\parallel \perp$ spectra).
The list of the observed bands in both polarizations is
reported in Table I. The number of observed phonon modes
is less than predicted by the selection
rules, but in any case the phonons detected in the HT phase are about
a half of those detected at LT, an obvious consequence of cell doubling.
\begin{table}[ht]
TABLE I. DMTTF-CA Raman active lattice \\ modes
in the LT and HT phases.
\begin{center}
\begin{tabular}{ccccc} \hline \hline
Mode no.~~~~& Raman, 45 K  	&~~~~~~& Raman, 110 K\\
\hline
$\nu_1$ & 36 ($\parallel \perp$)    &~~~~~~&  --  \\
$\nu_2$ & 42 ($\parallel \perp$)    &~~~~~~&   42 ($\parallel \perp$) \\
$\nu_3$ & 52 ($\perp \perp$) &~~~~~~&   --  \\
$\nu_4$ & 66 ($\perp \perp$) &~~~~~~&   62 ($\perp \perp$) \\
$\nu_5$ & 70 ($\perp \perp$) &~~~~~~&   67 ($\perp \perp$) \\
$\nu_6$ & 77 ($\perp \perp$) &~~~~~~&   --  \\
$\nu_7$ & 88 ($\parallel \perp$)&~~~~~~&   82 ($\parallel \perp$) \\ \hline \hline
\end{tabular}
\end{center}
\end{table}
%
The temperature dependence of the Raman frequencies are shown in the bottom
panel of Fig.~\ref{fig:peierls}.
One immediately notice the usual softening for all the modes as
we increase the temperature, due to lattice expansion. However, in the LT phase the
softening of some phonons is more pronounced close to the critical temperature. The frequency lowering is not as
large as in the HT phase (top panel of Fig.~\ref{fig:peierls}),
but is certainly present. The relative weakness of the effect
compared to the HT phase can be explained considering that in the HT
phase we observe the softening of an ``effective'' Peierls mode,
superposition of several phonons all coupled to the CT electrons.
In the LT phase, on the other hand, the softening is distributed
on several modes, and the phonon description and mixing changes
as we approach the phase transition. The effect is clearly seen in
Fig.~\ref{fig:peierls}, where at about 40 K there is a case of avoided crossing of
two phonons located around 70 \cm. In addition, we have
to keep in mind that the DMTTF-CA transition is second order,
but cannot be considered a strictly one-dimensional Peierls transition,
as the transition implies a change in the number of stacks per
unit cell. The actual phase transition mechanism is a complex one,
as shown by the fact that just below $T_c$ there is coexistence of two
slightly different degrees of ionicity on the molecular sites
(Section \ref{sec:valence}). This finding might be explained
in terms of an inequivalence
of the two stacks inside the unit cell in proximity of $T_c$.

\section{Discussion and conclusions}
\label{sec:disconcl}

The present work does not allow us to draw definitive conclusions
about the equivalence/inequivalence of the two DMTTF-CA
stacks inside the unit cell of the LT phase (anti-ferroelectric
or ferrielectric arrangement). As just discussed,
the presence of two slightly different degrees of ionicity
may imply that just after the phase transition we have a temperature
interval ($\sim 62-54$ K) in which the two stacks are inequivalent
with the $P1$ structure,\cite{collet01}, followed
by a definitive structural rearrangement yielding to the
14 K $P\bar 1$ structure.\cite{horiuchi03}
This picture would support the $P1$
structural determination,\cite{collet01} collected at 40 K below our coexistence $T$ region, only assuming that
the $P1$ structure refers to a non-equilibrium phase.
Such ``frozening'' of a metastable phase may be a consequence
of a too fast sample cooling, a case not
uncommon in organic solid state, although most of the times
refers to some disordered, glassy phase.\cite{kwok90}

At this point we wish to underline that inequivalence of the stacks
does not necessarily imply
an appreciably different degree of ionicity. DFT calculations
made for the $P1$ structure at 40 K indeed found practically identical
$\varrho$ for the two stacks with different dimerization
amplitudes.\cite{oison04} 
We may then have a scenario with two (slightly)
inequivalent stacks, but practically identical $\varrho$.

IR spectra
polarized parallel to the stack (Fig.~\ref{fig:reflpar}) of
course cannot disentangle dimerization amplitudes on different
stacks. Optical spectroscopy selection rules, on the other hand,
are based on the factor group (unit cell group), therefore reflecting
the translational long-range order of the crystal.\cite{walmsley} From this perspective,
two findings are in favor of inequivalent stacks, down to at
least 20 K. The first fact refers to the Raman-IR coincidence
observed for the $a_g$ molecular modes in the LT phase (Fig.~\ref{fig:RamanIR}).
If the two stacks are equivalent, and connected by an inversion center,
each $a_g$ mode of one stack would be coupled in-phase and out-of-phase
with the same mode on the other stack. The in-phase mode is Raman active, and
the out-of-phase one IR active, therefore we should not observe
precise frequency coincidence, the difference being related
to the strength of inter-stack interaction. Unfortunately, our
data are not conclusive in this respect, due to the frequency uncertainties
associated with the Kramers-Kronig transformation.\cite{note1} 

The other experimental observation that can be explained
in terms of inequivalent stacks is the doubling of
localized electronic transitions below 65 K.\cite{aoki93}
This experimental observation was the first one that
induced Aoki \textit{et al.}\cite{aoki93} to suggest the coexistence
of neutral and ionic species, but since then it has been
almost forgotten. Horiuchi \textit{et al.},\cite{horiuchi01}
as well as the present measurements, exclude such coexistence,
and the only explanation we can think of the doubling is
in term of ordinary Davydov splitting.\cite{walmsley}
However, the two components of the Davydov splitting
can be both optically active only in the lack of inversion
center relating the two stacks, otherwise the \textit{gerade}
component is inactive.

Only the replica of structural measurements and/or of the
refinement process starting from the two different
hypothesis will definitely settle the question
of equivalence-inequivalence of DMTTF-CA stacks.
On the other hand, this question is not particularly
relevant as far as the mechanism of DMTTF-CA
phase transition is concerned. The present analysis
departs from the previous ones\cite{horiuchi01,horiuchi03,okimoto05}
only in some seemingly
marginal details, but actually the resulting picture of the
phase transition is completely different. 

First of all, we have ascertained that the phase
transition implies only a limited change of $\varrho$,
DMTTF-CA remaining on the \textit{neutral} side down to
the lowest temperature. Furthermore, the
major charge rearrangement
\textit{follows}, by a few degrees K, the onset of cell doubling
and stack dimerization. The latter finding can be well appreciated from
Fig. \ref{fig:ionint}, where the inflection of the $\varrho(T)$ curve
occurs around 61 K, rather than at 65 K for the symmetry breaking.
Close scrutiny of Fig. 5
of Ref.~\onlinecite{horiuchi01} conveys the same information.
Therefore the transition can hardly be termed N-I, since cell
doubling and stack dimerization clearly constitute the driving force
of the transition. Indeed, we can envision a scenario in which
the dimerization and cell doubling lead to a better molecular
packing, with an increase in the Madelung energy and consequent
small, continuous change in $\varrho$. The presence of slightly differently
charged molecular species before the dimerization/cell doubling
has reached completion (Fig.~\ref{fig:ionint}) fits quite naturally into this
picture.

Disentangling the contribution of the cell doubling from that of the
stack dimerization is a useless endeavor. In any case, our measurements have clearly
evidenced the presence of an effective soft mode
along the chain (Fig.~\ref{fig:peierls}, top panel), so a Peierls-like mechanism is
certainly at work  in the precursor regime of the phase transition.
X-ray diffuse scattering also reveals the importance of electron-phonon
coupling along the chain, and of one-dimensional correlations.
It has been interpreted in terms of lattice relaxed
exciton strings (LR-CT) rather than in terms of a soft mode.
The phase transition is then regarded more as a disorder-order transition
(ordering of LR-CT exciton strings along and across the chains), and not as a displacive
one, with progressive uniform softening of the Peierls mode up to the final chain
dimerization. It would be interesting to re-analyze the X-ray
diffuse scattering data to examine whether 
and to what extent they are compatible with the soft-mode picture.
The present results of course only evidence the soft-mode
mechanism, and the presence of LR-CT exciton strings cannot be excluded, in particular
close to $T_c$, in the region where our data cannot be
unambiguously interpreted.

The LR-CT exciton string picture has been invoked
mainly to account for the dielectric constant anomaly
at $T_c$, attributed to the progressive ordering of the
a para-electric phase before reaching anti-ferroelectric (or ferrielectric) ordering.\cite{horiuchi03,collet02} 
On the other hand, it has been shown that the Peierls
mechanism is also able to \textit{quantitatively} explain the experimental
increase of the dielectric constant at $T_c$, interpreted
as due to charge oscillations induced by the Peierls mode.\cite{soos04}
We underline in this respect that although the dimerization
of the stack in the I phase has been often attributed to a spin-Peierls
mechanism,\cite{horiuchirev} the electronic degrees of freedom
are involved as well, particularly in proximity of the N-I
borderline.\cite{soos04} In addition, the dimerization transition
may occur also on the N side, provided the electron-phonon
interaction is strong enough. This is just the present case, and
correspondingly we have an increase of the dielectric constant
less important than in the case of TTF-CA,\cite{girlando04,horiuchirev} 
as predicted by the calculations.\cite{soos04}
In summary, the present interpretation stresses the importance
of the lattice instability over that of charge instability
in DMTTF-CA and related compounds.

\section{Acknowledgments}
We gratefully thank N. Karl for providing the DMTTF-CA crystals.
The reflectivity data have been fitted by the freely
available RefFit program (optics.unige.ch/alexey/reffitt.html).
We thank A. Painelli for many useful discussions.
Work in Italy supported by the ``Ministero dell' Universit\`a
e Ricerca'' (MUR), through FIRB-RBNE01P4JF and PRIN2004033197\_002.


\begin{thebibliography}{00}
\bibitem{note1}
See discussion
in Section \ref{sec:valence}. For parallel polarization we do not have the
check by absorbance, as the sample is completey opaque in the parallel
polarization.
\end{thebibliography}
\end{document}